\documentclass[]{aa}  

\usepackage[]{natbib}
\usepackage{graphicx}
\usepackage{txfonts}
\usepackage{hyperref}
\begin{document}

   \title{Mapping PAH sizes in NGC 7023 with SOFIA}
        
   \author{B.A. Croiset \inst{1}
          \and A. Candian  \inst{1}
          \and O. Bern\'{e} \inst{2}$^,$ \inst{3}
          \and A. G. G. M. Tielens \inst{1}}
 
   \institute{Leiden Observatory, Leiden University, Niels Bohrweg 2, 
                                2333 CA Leiden, The Netherlands              
           \and Universit\'{e} de Toulouse, UPS-OMP, IRAP, Toulouse, France
           \and CNRS, IRAP, 9 Av. colonel Roche, BP 44346, 31028 Toulouse Cedex \\ \email{croiset@strw.leidenuniv.nl}}

   \date{Received 9 November 2015 / Accepted 29 February 2016}

 \abstract
{ NGC 7023 is a well-studied reflection nebula, which shows strong emission from polycyclic aromatic hydrocarbon (PAH) molecules in the form of aromatic infrared bands (AIBs). The spectral variations of the AIBs in this region are connected to the chemical evolution of the PAH molecules which, in turn, depends on the local physical conditions.}
{Our goal is to map PAH sizes in NGC 7023 with respect to the location of the star. We focus on the North West (NW) photo-dissociation region (PDR) and the South PDR of NGC 7023 to understand the photochemical evolution of PAHs, using size as a proxy.}
 {We use the unique capabilities of the Stratospheric Observatory for Infrared Astronomy (SOFIA) to observe a 3.2' x 3.4' region of NGC 7023 at wavelengths that we observe with high spatial resolution (2.7") at 3.3 and 11.2 $\mu$m. We compare the SOFIA images with existing images of the PAH emission at 8.0 $\mu$m (\textit{Spitzer}), emission from evaporating very small grains (eVSG) extracted from \textit{Spitzer}-IRS spectral cubes, the extended red emission (Hubble Space Telescope and Canadian French Hawaiian Telescope), and H$_2$ (2.12 $\mu$m). We create maps of the 11.2/3.3 $\mu$m ratio to probe the morphology of the PAH size distribution and the 8.0/11.2 $\mu$m ratio to probe the PAH ionization. We make use of an emission model and of vibrational spectra from the NASA Ames PAH database to translate the 11.2/3.3 $\mu$m ratio to PAH sizes.}
 {The 11.2/3.3 $\mu$m ratio map shows the smallest PAH concentrate on the PDR surface (H$_2$ and extended red emission) in the NW and South PDR. We estimated that PAHs in the NW PDR bear, on average, a number of carbon atoms (N$_\mathrm{c}$) of $\sim$ 70 in the PDR cavity and $\sim$ 50 at the PDR surface. In the entire nebula, the results reveal a factor of 2 variation in the size of the PAH. We relate these size variations to several models for the evolution of the PAH families when they traverse from the molecular cloud to the PDR.}
{The high-resolution PAH size map enables us to follow the photochemical evolution of PAHs in NGC 7023.  Small PAHs result from the photo-evaporation of VSGs as they reach the PDR surface. Inside the PDR cavity, the PAH abundance drops as the smallest PAH are broken down. The average PAH size increases in the cavity where only the largest species survive or are converted into C$_{60}$ by photochemical processing.} 

   \keywords{ Astrochemistry -- photo-dissociation region (PDR) -- infrared: ISM --  ISM: molecules -- NGC 7023 -- SOFIA }

   \maketitle
%
%
\section{Introduction}
The aromatic infrared bands (AIBs) are prominent emission bands that are emitted by UV-irradiated interstellar matter. The strongest AIBs are observed at 3.3, 6.2, 7.7, 8.6, 11.2 $\mu$m and these features are produced by a specific class of molecules: Polycyclic Aromatic Hydrocarbons (PAHs) \citep{Sellgren1984,Leger1984,Allamandola1985,Puget1989}. PAHs are stochastically heated by UV radiation and become highly vibrationally excited; then they relax by emitting photons in the AIBs (for a recent review see \cite{Tielens2008}). The AIBs show systematic variations in relative intensity and central wavelength of the bands \citep{Peeters2002}, which reflect the local physical conditions and the chemical evolution of this species. \\
\indent NGC 7023, also known as the Iris nebula, is a well studied reflection nebula in Cepheus. It is irradiated by a triple binary system, HD 200775, consisting of two Herbig Be stars (T$_{\mathrm{eff}}$ $\sim 19000$ K) and a third poorly studied companion \citep{Li1994,Benisty2013}. The stars are located $320\pm51 $ pc from the Sun \citep{Benisty2013}. In this study we focus on two photo dissociation regions (PDRs) near the star system, one in the north west (NW) and one to the south. Both PDRs show strong emission from PAHs and very small grains (VSGs); in the NW PDR the presence of the neutral and cationic forms of the fullerene C$_{60}$ are also detected \citep{Berne2007,Berne2008,Sellgren2007,Sellgren2010,Berne2013}.\\
\indent The Stratospheric Observatory For Infrared Astronomy (SOFIA) is a telescope with an effective diameter of 2.5 m that is carried by a Boeing 747 aircraft \citep{Young2012}. SOFIA observes between 37,000 and 45,000 feet and above 99\% of the water vapor in the atmosphere. SOFIA can observe in the wavelength range of 0.3 - 1600 $\mu$m and has spectroscopic and imaging capabilities, which makes it a perfect tool to study the interstellar medium.\\
\indent Two of the prominent AIBs are due to C-H modes in PAHs, the 3.3 $\mu$m band (C-H stretch) and the 11.2 $\mu$m band (solo out of plane bending mode) \citep{Leger1984}. As both bands originate in neutral PAHs  \citep{Langhoff1996,Allamandola1999} and they span a wide wavelength range, the 11.2/3.3 $\mu$m ratio is a good measure of the sizes of PAHs \citep{Allamandola1985,Allamandola1989}. The 6.2 (and 7.7) $\mu$m bands are prominent in ionized PAHs and the 6.2/11.2 $\mu$m ratio is a good measure of the degree of ionization of the PAHs \citep{Peeters2002,Galliano2008}. Here, we  use the 11.2/3.3 $\mu$m ratio to study size variations in the PDR of NGC 7023 and relate the size to other characteristics of the interstellar PAH population and to the physical conditions in these PDRs.\\
\indent The outline of this work is as follows:  observations and  data reduction are discussed in Section 2. In Section 3, the maps are presented. Section 4 presents a detailed description of how the 11.2/3.3 $\mu$m ratio is translated into a PAH-size model. The results are discussed in Section 5 and conclusions are drawn in Section 6.

%
%

\section{Observations and data reduction}
NGC 7023 was observed by SOFIA in the $3^{\mathrm{rd}}$ cycle with two instruments: the First Light Infrared TEst CAMera (FLITECAM) and the Faint Object infraRed CAmera for the Sofia Telescope (FORCAST) \citep{Temi2012,Herter2012}. \\
\indent FLITECAM is an infrared camera that operates in the  wavelength range of $1.0 - 5.5\ \mu$m. The instrument was used in nod mode with a nod throw of 600 arcsec and in a 3-point dither pattern with a dither offset of 10 arcsec and 100s exposure per dither. The filter used for the observations that we report is the PAH-filter, which has a central wavelength ($\lambda_c$) of 3.30 $\mu$m and a very narrow bandwidth ($\Delta \lambda$) of $0.09\  \mu$m. This filter is small enough to only trace the 3.3 $\mu$m PAH emission. The instrument covers a region of 8 arcmin $\times$ 8 arcmin with a square plate scale of 0.475 arcsec on a pixel array of 1024 $\times$ 1024 pixels.\\
\indent The FORCAST instrument is a dual-channel mid-infrared camera and spectrographically sensitive in the wavelength range of 5-40 $\mu$m. In this observation, the Short Wave Camera (SWC) was used with the 11.1 $\mu$m filter, where $\lambda_c =$ 11.1 $\mu$m and $\Delta \lambda  =  0.95\ \mu$m. For the observations, the nod match chop mode (C2N) was used with a chop throw of 300 arcsec, a chop angle of 15 degrees, and no dithering. The SWC has 256 $\times$ 256 pixel array, which yields a field of view (FOV) of 3.2 arcmin $\times$ 3.4 arcmin with a square plate scale of 0.768 arcsec.
\begin{figure}[ht]
 \centering
 \includegraphics[width=1\linewidth]{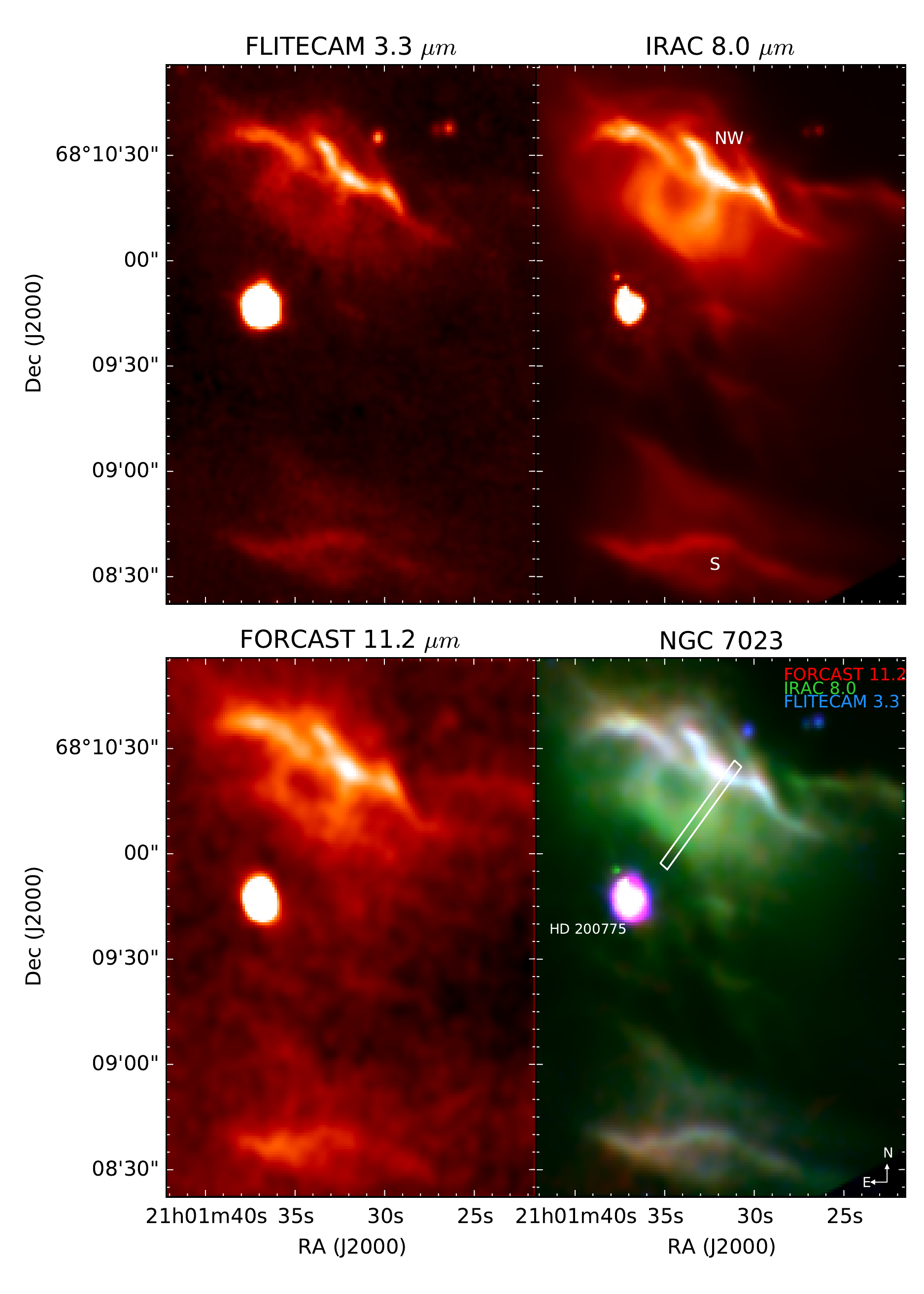}
 \caption{Top left panel: FLITECAM 3.3 $\mu$m PAH filter observation of NGC 7023 with a $\Delta \lambda = 0.09\ \mu$m. The top right panel: IRAC 8.0 $\mu$m filter with a $\Delta \lambda = 2.90\ \mu$m. The NW and the S indicate the NW PDR and the S PDR. The bottom left panel: FORCAST 11.1 $\mu$m SWC filter observation with a $\Delta \lambda = 0.95\ \mu$m. The bottom right panel is an RGB image of the three images with (in red) the FORCAST, the IRAC (in green), and   the FLITECAM image (in blue). All images are normalized to the maximum value in the north PDR. The white box is the region used in the crosscut (see Fig.~\ref{fig:cutout}).} 
 \label{fig:observations}
\end{figure}  \\
\indent The exposure times for FLITECAM were 300 s on source and 660 s total. For FORCAST the exposure times were 2300 s on source and a total overhead of 3557 s. 
The FLITECAM instrument has a resolution of $\sim$ 2 arcsec, which is limited by telescope jitter. FORCAST suffers slightly more from telescope jitter and has a resolution of 2.7 arcsec.\\
\indent In addition to SOFIA observations, we used  images provided by the IRAC instrument of the \textit{Spitzer} space telescope \citep{Werner2004} to analyze the nebula. The IRAC 8.0 $\mu$m filter was used, which has a $\lambda_c$ of 8.0 $\mu$m and a broad bandwidth of 2.9 $\mu$m, thus it includes the strong PAH features at 6.2, 7.7, and 8.6 $\mu$m.
We also used images of extended red emission (ERE), VSG emission and H$_2$ emission from different telescopes. With the \textit{Spitzer} IRS instrument, the emission of VSGs and neutral and ionized PAH were extracted with the blind signal separation (BSS) method by \cite{Berne2007}. Molecular hydrogen was observed with the Perkins Telescope in the S PDR and with the Canadian French Hawaian Telescope (CFHT), using the 2.12 $\mu$m filter in the NW PDR \citep{Lemaire1996,Sellgren2003}. With observations from the latter telescope an ERE image of the South PDR was extracted by \cite{Berne2008}. For the NW PDR an ERE image was made from Hubble Space Telescope images (see \cite{Berne2008}). \\
\indent Fig. \ref{fig:observations} presents the images of NGC 7023 observed with FLITECAM at 3.3 $\mu$m, IRAC at 8.0 $\mu$m, FORCAST at 11.2 $\mu$m, and a combined RGB image. The images show the NW PDR, the South PDR, and the triple binary system. The Herbig Be stars are unresolved and located in the east middle of the image ($\alpha: 21^h\ 01^m\ 36.92^s,\ \delta: +\ 68^d\ 09^m\ 47.76^s $). The 8.0 $\mu$m \textit{Spitzer} image most likely resolves the outermost companion that is located 6 arcsec away from the central star \citep{Li1994}. In the north west quadrant image, three stars are located. The stars are brighter at 3.3 $\mu$m than at 8.0 $\mu$m and absent at 11.2 $\mu$m. 
\begin{figure*}[t]
  \includegraphics[width=1\linewidth  ]{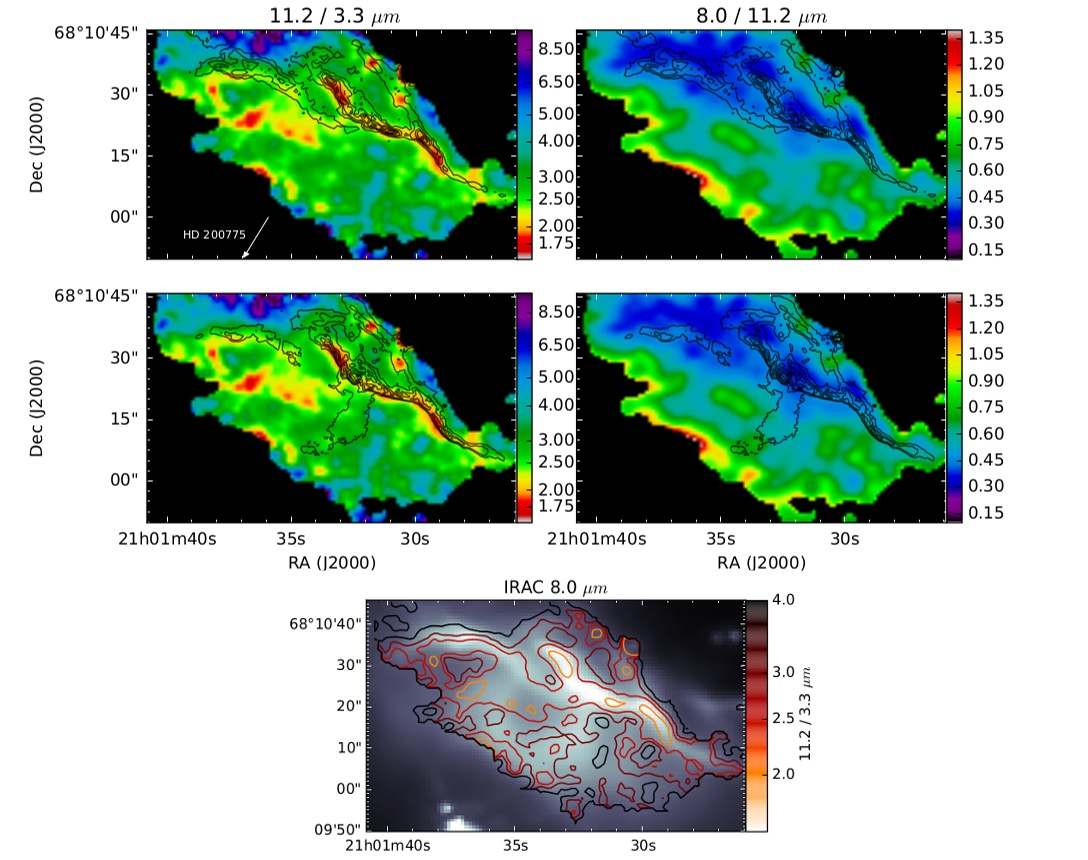}
 \caption{PAH ratio maps of NGC 7023 NW PDR. The colors correspond to the 11.2/3.3 $\mu$m emission ratio on the left (PAH size tracer) and to the 8.0/11.2 $\mu$m ratio on the right (PAH ionization indication). Top panels: The contour lines represent the H$_2$ emission \citep{Lemaire1996} where the contours range from 0.8 to 3.7 [Wm$^{-2}$sr$^{-1}$] with 5 steps. Middle panels: The contour lines are the ERE emission. The contours are spaced in 5 steps with the outermost contour line of 0.5 and a maximum of 1.0 (Arb. Units) \citep{Berne2008}. Pixels with an S/N smaller than 3 in the ratio have been discarded. Bottom panel: The IRAC 8.0 $\mu$m image with the contours of the 11.2/3.3 $\mu$m ratio. }
 \label{fig:PDR_NW}
\end{figure*}\\
\indent  The images at 3.3, 8.0, and 11.2 $\mu$m have a signal-to-noise
(S/N) of 76, 45, and 21 in the peak of the NW PDR, respectively. The 11.2 $\mu$m image was convolved with a 2 pixel (1.4 arcsec) Gaussian kernel to increase the S/N. The  field of view (FOV) of the images is $\sim 1.9$ arcmin $\times$ 2.5 arcmin. The raw images from SOFIA were rotated and misaligned with respect to the 8.0 $\mu$m image. Stars outside the images, shown in Fig.~\ref{fig:observations}, and the three stars were used as reference to realign the images to the coordinates of the \textit{Spitzer} image with high accuracy, using an optimizing algorithm. However, at 11.2 $\mu$m the stars were not detected, therefore HD 200775, the NW, and South PDR were used to align the images. The uncertainty on the alignment is less than 1 arcsec. 
The 3.3 $\mu$m and the IRAC image were regridded to the same pixel scale of the FORCAST image (0.768 arcsec). 
Using the aligned images we  made intensity ratio maps of the nebula (see Fig.~\ref{fig:PDR_NW}, \ref{fig:PDR_S}). \\
\indent To be able to only take the PAH emission into account, we estimated the continuum contribution in the 3.3 $\mu$m and the 11.2 $\mu$m filters. The contribution was \textasciitilde10-20\% in the 3.3 $\mu$m filter observed with AKARI by \cite{Pilleri2015} and $\sim 20$\% when determined with the \textit{Spitzer} IRS spectra for the same positions observed by AKARI.
 
%
%

\section{Morphology of the maps}

NGC 7023 NW is often considered as the prototype of a PDR since it has an edge-on structure, which shows a clear stratification of  different species \citep{Pilleri2012}. The star has created a low-density cavity that is surrounded by a dense molecular cloud. The cavity edge is lined by a thin HI layer \citep{Fuente1998}. The fullerene abundance increases rapidly into the cavity, while the PAH abundance decreases \citep{Sellgren2010, Berne2012, Pilleri2012}.\\
\begin{figure}[ht]
 \includegraphics[width=  \linewidth]{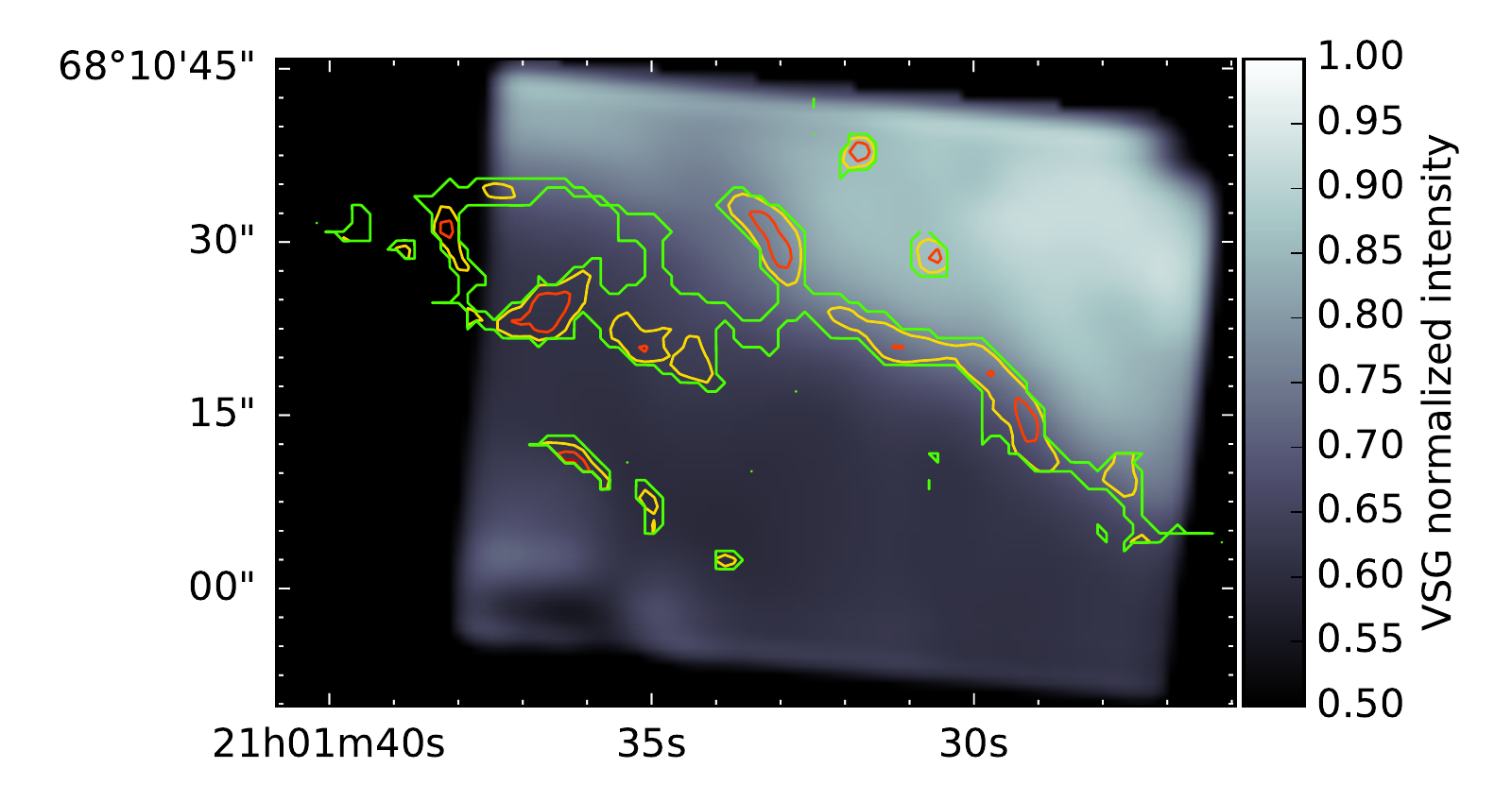}
 \caption{VSG emission extracted from \textit{Spitzer} IRS by \cite{Berne2007}. The contours are the 11.2/3.3 $\mu$m ratio lower than 2.4 (Fig.~\ref{fig:PDR_NW}). The values 1.9, 2.1, 2.4 of the 11.2/3.3 $\mu$m ratio contours are drawn.} 
  \label{fig:VSGmap} 
\end{figure} 
\indent In Fig. \ref{fig:observations}, we observe the bright Northern PDR surface which is shaped in a wall structure, which  separates the molecular cloud and the PDR cavity. Strong PAH emission is observed from the PDR surface and extends to the south-east to 18 arcsec. The PDR surface is quite apparent in all wavelengths and coincides well with the tracers of the PDR surface, such as H$_2$ and the ERE \citep{Lemaire1996,Berne2008}. We note the presence of a ring-like structure in the source south of the NW PDR.
South of the stars, the cavity is larger and faint extended emission appears, most prominent at 8.0 $\mu$m. At 11.2 $\mu$m the extended emission is not clearly visible owing to higher noise levels. The South PDR surface has a similar wall structure that is clearly visible in the images.\\
\indent From the observed images, two maps are created: the 11.2/3.3 $\mu$m ratio and 8.0/11.2 $\mu$m ratio. A zoomed-in version of the NW PDR is shown in Fig.~\ref{fig:PDR_NW} with contours of H$_2$ emission (2.12 $\mu$m) and ERE, and a zoomed-in version of the South PDR in Fig.~\ref{fig:PDR_S} with contours of ERE and H$_2$ emission. There is a very clear alignment between the PDR surface that is located 48 arcsec away from the star traced by H$_2$ and the 11.2/3.3 $\mu$m emission ratio. The intensity ratio is higher in the PDR surface and rises quite rapidly behind the PDR and away from the star into the molecular cloud. The ring structure below the NW PDR does not appear in the ratio maps, nor in the H$_2$ image, and is only partly visible in the ERE image. We note that the VSGs emission is limited to the molecular cloud region (see Fig.~\ref{fig:VSGmap}). The minimum in the 11.2/3.3 $\mu$m ratio also indicates the boundary of the emission of the VSGs. \\
\indent A similar good alignment is observed in the South PDR with the location of the PDR surface, as traced by the ERE (HST), H$_2,$ and 11.2/3.3 $\mu$m ratio. A red spot is apparent between the PDR surface and HD 200775 and is due to a star behind the reflection nebula. The star is fainter than the three stars above the NW PDR (see Fig.~\ref{fig:observations}) in the 2MASS observation \citep{2MASS}.
In the 8.0/11.2 $\mu$m ratio maps, the emission decreases with distance from the star and has a maximum at the cavity edge. The PDR surface is located in an area where the ratio is low and the intensity ratio drops (dark blue) right behind the PDR surface and into the molecular cloud.\\
\indent In Fig.~\ref{fig:PDR_S}, a spatial map of the 11.2/3.3 $\mu$m ratio and 8.0/11.2 $\mu$m ratio of PAH in the South PDR is shown. These maps are made in the same way as for the NW PDR, but the ERE image and H$_2$ are made from different telescopes \citep{Sellgren2003,Berne2008}. The main PDR surface is traced by ERE, H$_2,$ and the 11.2/3.3 $\mu$m ratio, however a second PDR surface to the south is not traced by the ERE emission. Compared to the NW PDR, the 11.2/3.3 $\mu$m ratio traces the PDR surfaces quite well, just like in the NW PDR, although the ratio is somewhat higher. 
Unfortunately the S/N of the 11.2 $\mu$m was not high enough to map the extended emission to the north of the South PDR (see the IRAC 8.0 $\mu$m panel in Fig.~\ref{fig:PDR_S}). In the 8.0/11.2 $\mu$m ratio map, the maximum value is lower than in the NW PDR and decreases with distance from the star.\\
\begin{figure*}[t]
  \includegraphics[width=1\linewidth]{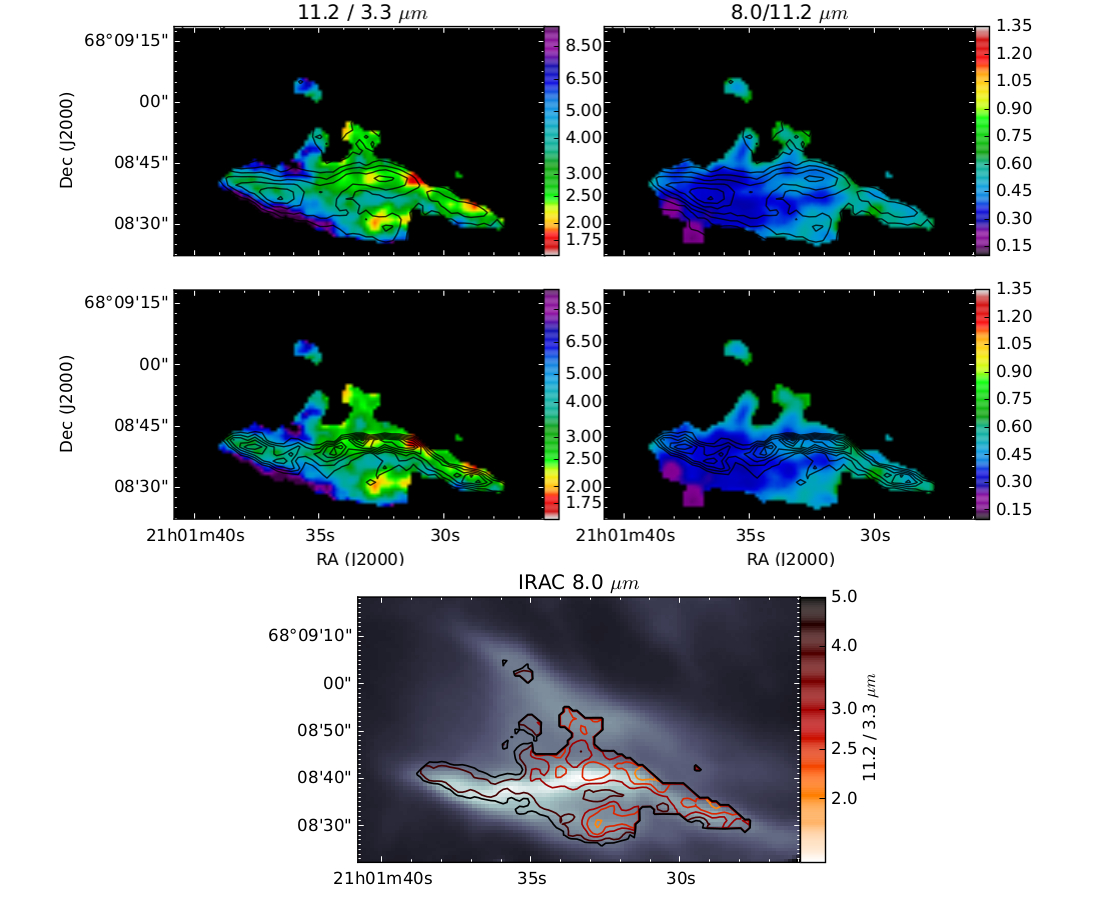} 
 \caption{Intensity ratio maps of the South PDR of NGC 7023. The colors from the colorbar correspond to the ratio of the 11.2/3.3 $\mu$m ratio on the left and the 8.0/11.2 $\mu$m ratio on the right. In the top panels the contour lines follow the H$_2$ emission (2.12 $\mu$m) from 90 to 150 in 6 steps (Arb. Units) \citep{Sellgren2003}. In the middle panels the contour lines are based on the ERE \citep{Berne2008}. The lines have been normalized and range from 0.1 to 1 (Arb. Units) with 6 steps. Pixels with a S/N smaller than 3 in the ratio have been discarded. Bottom panel: The IRAC 8.0 $\mu$m image with the contours of the 11.2/3.3 $\mu$m ratio.}
 \label{fig:PDR_S}
\end{figure*}

\section{Results}
\subsection{PAH size-emission model}
\begin{figure}[t]
  \centering
 \includegraphics[width=\linewidth]{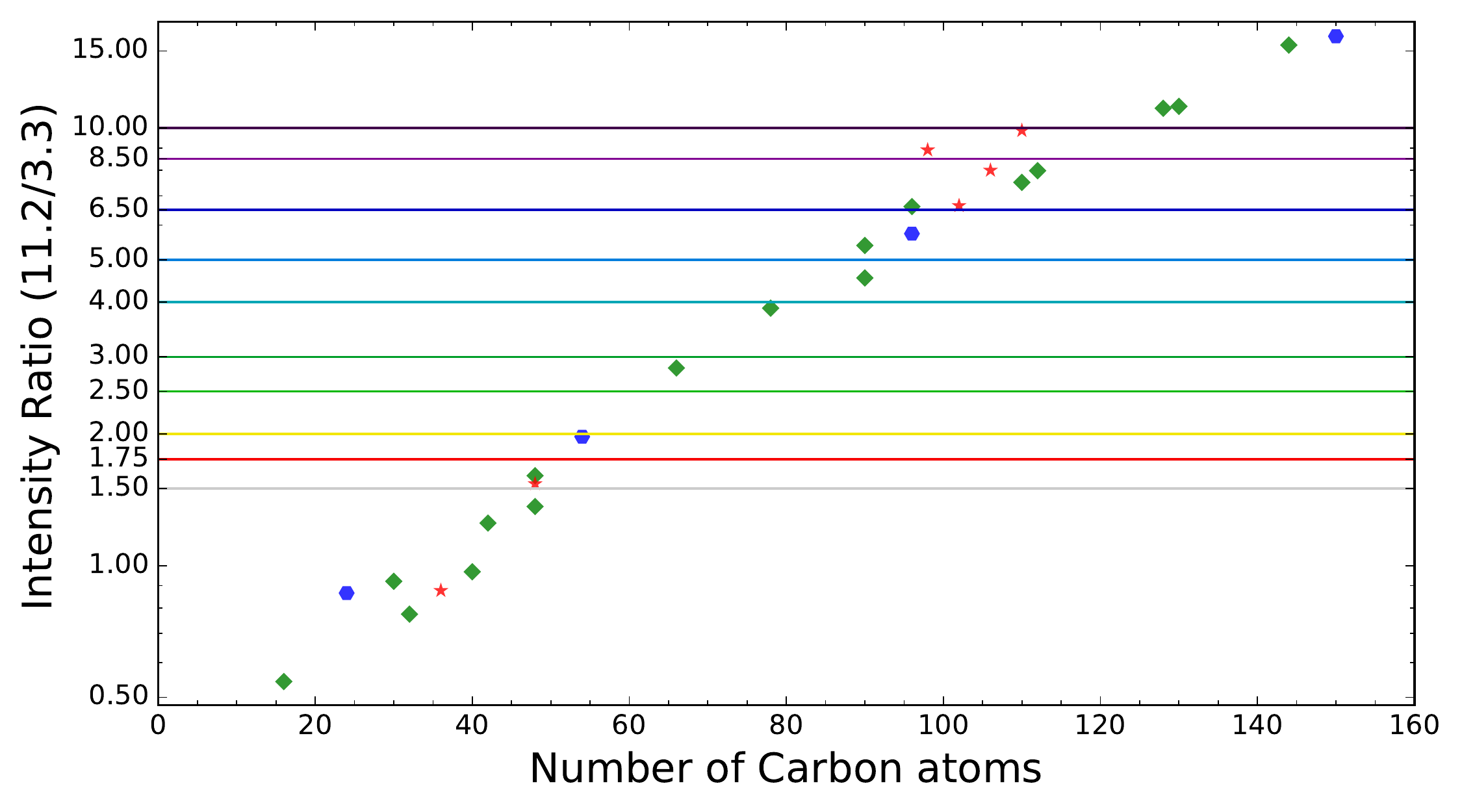}
\caption{11.2/3.3 $\mu$m intensity ratio of 27 different PAHs with blue coronene family, green compact, and red non-compact PAHs. The average photon energy absorbed by the PAHs used is 6.5 eV (see Eq.~\ref{photon}). The ratio of the C-H stretching mode (3.3 $\mu$m) over the solo out-of-plane bending mode (11.2 $\mu$m) is extracted from the theoretical spectra from the NASA Ames PAH database. The y-axis is linearly proportional to the adopted intrinsic band strength of the 11.2/3.3 $\mu$m ratio. The colored horizontal lines have the same color coding as used for the 11.2/3.3 $\mu$m ratio maps in Figs.~\ref{fig:PDR_NW} and \ref{fig:PDR_S}.}
 \label{fig:PAHsize}
\end{figure}

The 11.2/3.3 $\mu$m ratio is a good measure of the size of the PAH \citep{Allamandola1985,Ricca2012}. We can use the calculated intrinsic strength of the CH modes in PAHs in the NASA Ames PAH database \citep{BoersmaPAH} to calculate the 11.2/3.3 $\mu$m intensity ratio, taking the IR cascade fully into account. The results are shown in Fig.~\ref{fig:PAHsize}.

To estimate the emission of PAHs in NGC 7023, we need to provide E$_\mathrm{abs}$ the average energy per photon absorbed by a PAH molecule,
\begin{equation}
\label{photon}
  E_\mathrm{abs} =  \frac{\int N_{\nu} h \nu \sigma_{PAH,\nu} \exp{(-\kappa_{\nu}N_H)}  \mathrm{d}\nu}{\int N_{\nu}\sigma_{PAH,\nu} \exp{(-\kappa_{\nu}N_H)}\mathrm{d}\nu}
,\end{equation}
where $N_{\nu} \mathrm{d}\nu= F_{\nu} / h\nu \mathrm{d}\nu$, $\nu$ is the frequency, $h$ the Planck constant, $N_{\nu}$ is the number of photons $,\ \sigma_{PAH,\nu}$ is the absorption cross section of PAHs, $\kappa_\nu $ the absorption cross-section per H nucleon and N$_{\mathrm{H}}$  the column density. We used a Kurucz model for the emission of the star with a T$_{\mathrm{eff}}$ 19000 K $\pm 2000$ K \citep{Kurucz1993,Benisty2013}. For the value $\kappa_{\nu}$, we used an extinction law with R$_V$ = 5.5 \citep{Draine2001} and a N$_{\mathrm{H}}$ of 1.5  $\times 10^{21} [$atoms cm$^{-2}]$. We used the absorption cross-section of circumcoronene (C$_{54}$H$_{18}$) from the Cagliari PAH database \citep{Malloci2007}. The effect of N$_{\mathrm{H}}$ on the PAH 11.2/3.3 $\mu$m ratio is minimal, as illustrated in Fig.~\ref{fig:Extinction}. The mean absorbed photon energy for PAH molecules in NGC 7023 was evaluated to be 6.50 eV. The effects of the uncertainty in the stellar T$_{\mathrm{eff}}$ and the dust extinction contribute to a $\pm 0.3$ eV. The $\pm 0.3$ eV uncertainty on the mean absorbed photon energy is not a statistical uncertainty, but rather a systematic uncertainty. \\
%
%
%
%
\label{families}
\indent Fig. \ref{fig:PAHsize} shows the results for 27 different PAHs from multiple families, with the constraints that N$_\mathrm{c}$ is smaller than 150 and that the molecule contains at least one solo C-H bond (responsible for the 11.2 $\mu$m emission). We used PAHs molecules from the Coronene, Ovalene, Anthracene, Tetracene families \citep{Ricca2012,Peeters2015}, and the Pyrene family. While most of these PAHs are very compact, we have also included a few non-compact ones,to investigate the effect of the molecular structure on the 3.3/11.2 $\mu$m ratio. All the PAHs in the selections are shown in Fig. \ref{fig:families}). As the UV absorption cross-sections for all PAHs are very similar, the uncertainty  produces a vertical shift in Fig.~\ref{fig:PAHsize}.\\
\indent Figure \ref{fig:PAHsize} shows that the intensity ratio is sensitive to the size of the PAHs \citep{Allamandola1985} and forms a well-defined relationship \citep{Ricca2012}. The slope and shape of the relation is not sensitive to the stellar input spectrum or the extinction in the PDR. However, it can be shifted vertically for different average photon energies or adopted intrinsic strength.
 
%
\subsection{PAH size}
With the model (Fig.~\ref{fig:PAHsize}), the 11.2/3.3 $\mu$m ratio can be converted to PAH sizes. The horizontal colored lines in the figure have the same color coding as the 11.2/3.3 $\mu$m intensity ratio in the maps. In the NW PDR (orange in Fig.~\ref{fig:PDR_NW}), PAHs have a typical size of 50 carbon atoms. Moving towards the star, the ratio drops (green) and  PAHs have average sizes of N$_\mathrm{c}$= 60 $\sim$ 70. In the South PDR the average sizes of the PAHs are slightly larger with respect to the NW PDR, N$_\mathrm{c}$ = 60  (green) in the PDR surface, and increase towards the left up to N$_\mathrm{c}$ = 80 (light blue) in Fig.~\ref{fig:PDR_S}. \\
The results show that the 11.2/3.3 $\mu$m ratio is mainly sensitive to size and has little dependence on molecular structure. Matrix isolation and gas phase experiments, as well as theoretical density functional theory (DFT) calculations show that the intrinsic band-strength ratio of the C-H stretch to the out-of-plane-bending mode (3.3/11.2 $\mu$m) is very constant between different PAHs \citep{Joblin1994,Langhoff1996}. However, the ratios derived by these different techniques are different. Further experiments are needed to settle this issue. However, we emphasize that a different adopted band-strength ratio will only produce a vertical shift in Fig.~\ref{fig:PAHsize}. An uncertainty in the adopted intrinsic strength translates to a systematic uncertainty in quantifying PAH size. If the adopted intrinsic strength is increased by a factor of 2, then a range of N$_\mathrm{c}$ = $50-105$ would change to a range of N$_\mathrm{c}$ = $35-80$.  The trends, however, remain. 

\begin{figure}[t]
 \includegraphics[width=\linewidth]{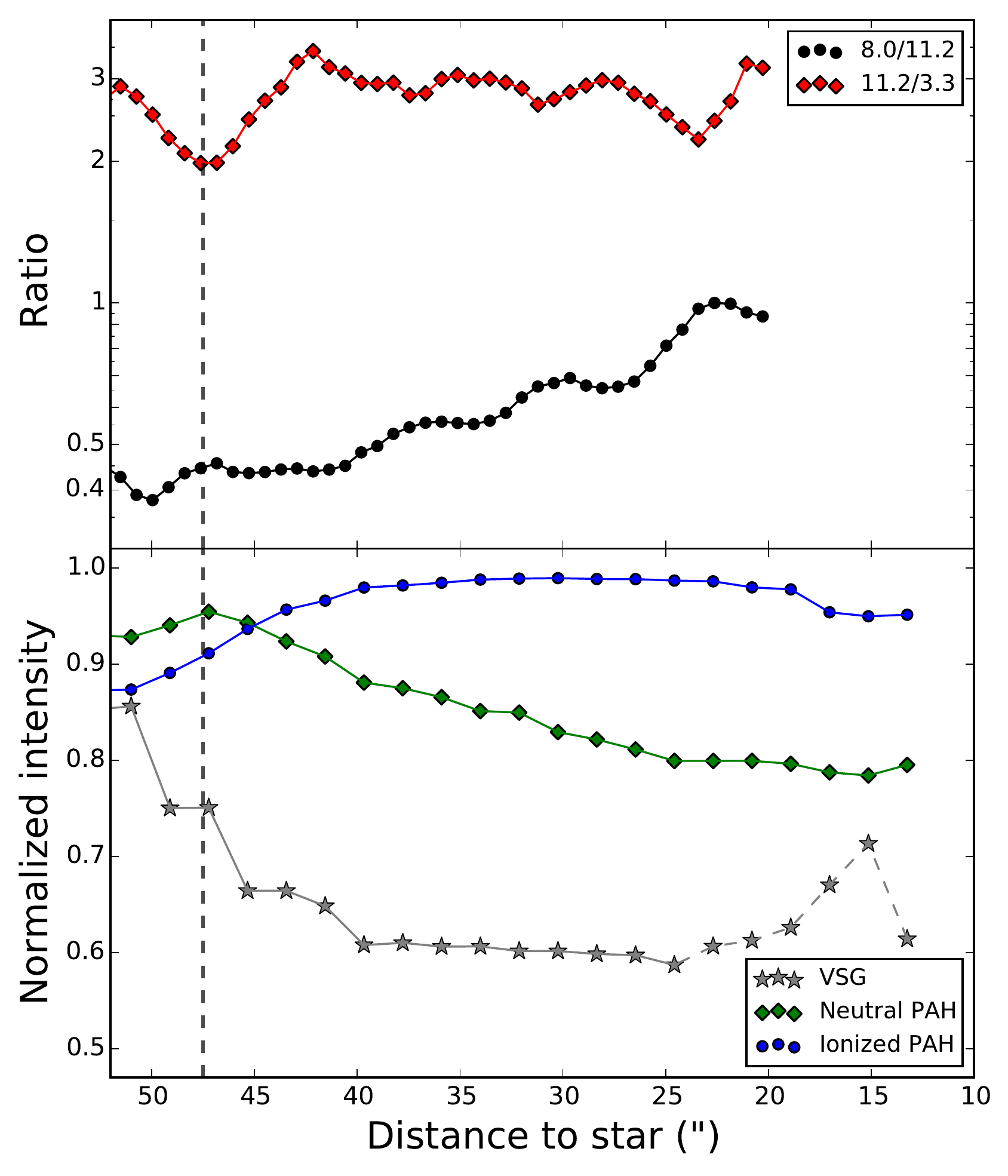}
  \caption{Top panel: The emission of the size (11.2/3.3 $\mu$m) and ionization map as a function of distance to the star in NGC 7023 (see Fig.~\ref{fig:PDR_NW}). The region of the crosscut is the white rectangle is Fig.~\ref{fig:observations}. In the bottom panel, along exactly the same region of the top panel, the normalized intensity of the neutral PAHs, ionised PAHs, and the VSGs is plotted \citep{Berne2007}. The vertical dashed line is the peak of the H$_2$ emission (see Fig.~\ref{fig:PDR_NW}). The PDR surface is 48 arcsec from the star, where the 11.2/3.3 $\mu$m ratio and the neutral PAH emission are maximum and the VSG emission starts to decline. The increase in VSG abundance for less then 25 arcsec (dashed) is an artifact and reflects emission from big grains in radiative equilibrium, rather than thermally fluctuating VSGs. }
 \label{fig:cutout}
\end{figure}

\subsection{PAH size variation in the PDR}
\indent Fig.~\ref{fig:cutout} shows a crosscut across the NW PDR in PAH sizes, ionization, neutral PAH emission, and VSG. The PDR surface is located 48 arcsec from the star. At this point the neutral PAH emission is the strongest and the 11.2/3.3 $\mu$m ratio the lowest, with sizes of N$_\mathrm{c}$ = 50. From the PDR surface to the cavity, the PAH sizes increase and remain constant with N$_\mathrm{c}$ $\simeq$ 70 and decrease in size just before the cavity to N$_\mathrm{c}$ $\simeq$ 60.  
The 8.0/11.2 $\mu$m ratio is enhanced owing to the steady drop in emission from neutral PAH in the 11.2 $\mu$m, while the broad 8.0 $\mu$m filter, which traces the ionized PAH emission bands, only has  a slight decrease in intensity towards the cavity. The 11.2/3.3 $\mu$m ratio remains constant in the PDR cavity, while the VSG emission drops 25\% through the PDR surface.

%

\section{Discussion}
The evolution of PAHs in the NW PDR of NGC 7023 has been studied by \cite{Berne2007,Pilleri2012,Boersma2013,Andrews2015} and modeled by \cite{Rapacioli2005,Berne2012,Montillaud2013,Berne2015}. This evolution is driven by the increasing strength of the radiation field as PAHs approach the surface of the PDR in the evaporative flow into the cavity. In this evolution, the weakest link goes first, i.e., bonds with the lowest binding energy will break first \citep{Jochims1996}.\\ 
\indent From these studies, the following picture emerges: in the molecular cloud, PAHs cluster into VSGs \citep{Rapacioli2005}, bound by typically 2.5 eV for 50 C-atom PAHs. This is much less than the measured binding energy of H-atoms or C$_2$H$_2$ groups \citep{Jochims1996}. Hence, when the VSGs approach the PDR surface, upon photon absorption, these clusters  evaporate into their constituent PAH molecules. As  observations show, at the PDR surface, most clusters have disappeared (Fig.~\ref{fig:VSGmap}). At that point, the next step in the photochemical evolution can ensue and photons will process these free-floating PAHs into their most stable isomers, which are likely the most compact PAHs (e.g., coronene and ovalene families). Inside the PDR cavity, these so-called grandPAHs \citep{Andrews2015} are then broken down, first through complete H-stripping and conversion into graphene flakes, followed by sequential steps of C$_2$ loss to smaller and smaller flakes, rings, and chains. The breakdown of the carbon skeleton competes with isomerization into cages and  fullerenes \citep{Berne2012}. The observations reveal a rapid drop in the abundance of PAHs and an increase in the abundance of C$_{60}$ (Fig.~2 in \citeauthor{Berne2012}, \citeyear{Berne2012}). We note that this destruction process commences when the PAH charge distribution shifts from neutrals towards cations. \\
\indent There is good experimental evidence for these photochemical steps in the evolution of PAHs \citep{Ekern1997,Joblin2009,Zhen2014a,Zheny2014b}. Recent experiments have demonstrated that, for large PAHs, ionization dominates over fragmentation \citep{Zhen2015}. For large PAHs, the first ionization potential is less than the average photon energy, so destruction is not expected to set in until the PAHs are ionized. \\
\indent The present observations provide further qualitative insight in this evolution and quantifies it in terms of the PAH sizes. The PAHs evaporating from VSGs deep in the PDR are relatively large (80 C-atoms) but photochemically unstable. When exposed to the harsh stellar radiation field at the surface of the PDR, these are rapidly converted to more stable and somewhat smaller (e.g., likely very compact) PAHs.  Inside the cavity, the abundance of PAHs drops and, as PAH destruction is a very strong function of size, the PAH size is expected to increase again, as our observations reveal, to $\sim$ 70 C-atoms (Fig.~\ref{fig:cutout}). These large PAHs can be photochemically converted into C$_{60}$ \citep{Zhen2014a,Berne2015}.

\section{Conclusions} 
Using FLITECAM and FORCAST on SOFIA, we have made a high spatial resolution map of the 11.2/3.3 and 8.0/11.2 $\mu$m intensity ratios to trace PAH size and ionization, respectively. The minimum of the 11.2/3.3 $\mu$m ratio (lower ratio for smaller PAHs) shows a remarkable alignment with the PDR surface traced by H$_2$ and ERE. The 8.0/11.2 $\mu$m ratio increases from the PDR surface to its maximum in the cavity, where the radiation field from the star ionizes and destroys the PAHs. A typical PAH size of 50 carbon atoms (circumcoronene) is found at the PDR surface. The analysis reveals a factor of 2 variation in the size of the emitting PAHs in the NW PDR in NGC 7023. This result is robust against variation in molecular structures. 
The PAH size distribution agrees with the distribution of \cite{Boersma2013}. The agreement between these very different approaches, 11.2/3.3 $\mu$m ratio versus fitting \textit{ Spitzer} IRS spectra with theoretical PAH spectra from the PAHdb, validates both techniques.\\
\indent The peak in the 11.2/3.3 $\mu$m ratio (corresponding to PAHs of  $\sim$ 50 C-atoms) also corresponds to the edge of the VSG emission  i.e., occurs at the transition between the molecular cloud and the PDR surface. From the PDR surface to the cavity, 11.2/3.3 $\mu$m ratio decreases, which suggests that the smallest PAHs are quickly destroyed by UV photons. Hence, the typical size increases to 70 carbon atoms. This is also in agreement with the idea that PAHs with sizes between 60 and 70 C-atoms can be converted into C$_{60}$ in these regions \citep{Zhen2014a,Berne2015}. \\
\indent This  behavior is consistent with the general model for the evolution of PAHs in the PDR of NGC 7023 developed by \cite{Rapacioli2005} and \cite{Pilleri2012} where the strong FUV flux from the central star first evaporates PAH clusters into their PAH constituents, converts the latter into the most stable ones, and then destroys them. Quantitatively, our observations reveal that this photochemical evolution is very size-dependent and most effective for PAHs containing $\sim$ 50 C-atoms , in good agreement with theoretical models \citep{Berne2015}. \\
\indent These results show that the capabilities of SOFIA can help reveal the general structure and PAH evolution in PDRs, complementing the observations performed with ISO and \textit{Spitzer}.

\begin{acknowledgements}
        We thank Kris Sellgren for her insightful comments and for sending us her H$_2$ images.
    The authors also thank Paolo Pilleri for useful comments and making his data available.
        B.A. Croiset acknowledges the SOFIA grant which allowed this research to be presented at the 30 years of PDRs symposium.
        Based on observations made with the NASA/DLR Stratospheric Observatory for Infrared Astronomy (SOFIA), SOFIA is jointly operated by the Universities Space Research Association, Inc. (USRA), under NASA contract NAS2-97001 and the Deutsches SOFIA Institut (DSI) under DLR contract 50 OK 0901 to the University of Stuttgart.
        This work was supported by the CNRS program Physique et Chimie du Milieu Interstellaire (PCMI). 
        Studies of interstellar PAHs at Leiden Observatory are supported through advanced European Research Council grant 246976
        and a Spinoza award.
        
\end{acknowledgements}

\appendix
\section{}
\begin{figure}[h]
   \includegraphics[width= \linewidth]{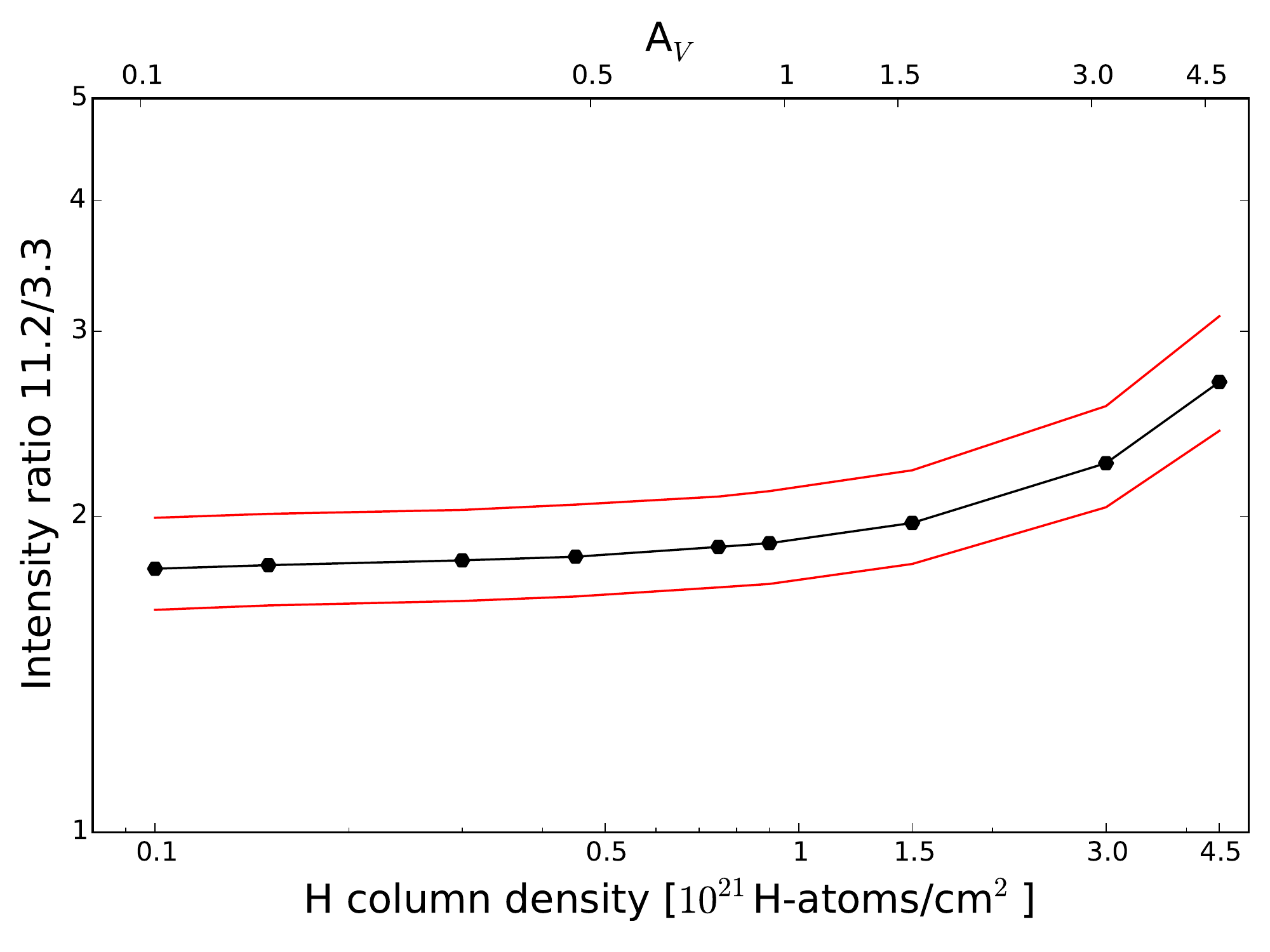}    
 \caption{Effect of dust extinction optical depth on the 11.2/3.3 $\mu$m ratio. The dust extinction cross-sections per H-atom have been taken from \cite{Draine2001}; the model with R$_{V}$=5.5). For comparison, the maximum H-column density along the line of sight in the northern PDR of NGC 7023 is measured to be 2.2 $\times 10^{21}$ atoms cm$^{-2}$ \citep{Fuente1998}.} 
\label{fig:Extinction}
\end{figure}

\begin{figure}[h]
  \includegraphics[width=  \linewidth]{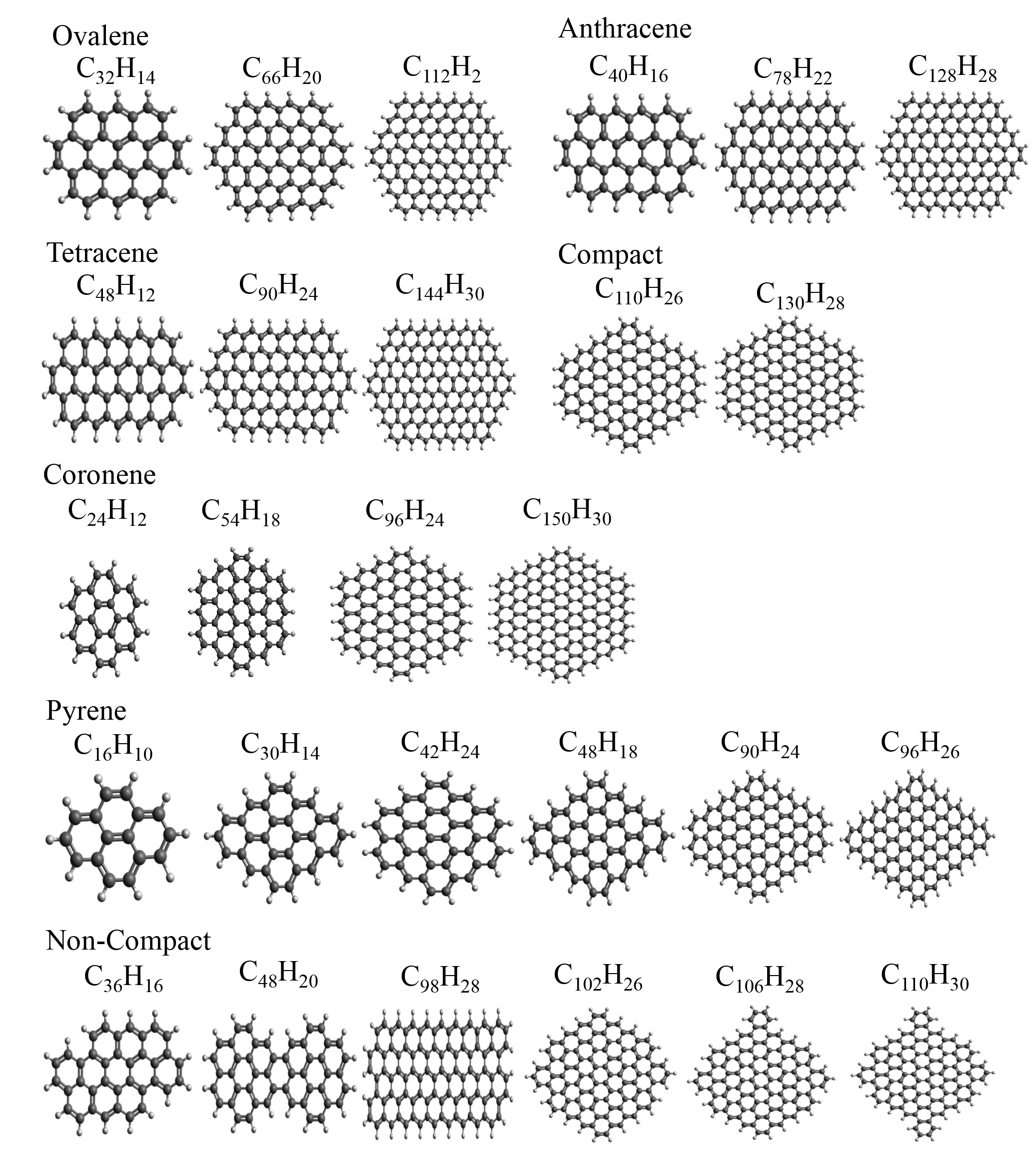}
 \caption{All 27 PAHs used for the model (see section \ref{families}).}
 \label{fig:families}
 \end{figure}

Fig.~\ref{fig:Extinction} shows the effect of visual extinction (with HI column density) on the 3.3/11.2 $\mu$m ratio. Fig.~\ref{fig:families} shows the sample PAHs used in this study.

\end{document}